\documentclass[review]{elsarticle}

\journal{Applied Thermal Engineering}

\usepackage{amsmath}

\usepackage{eurosym}

\usepackage{subcaption}

\usepackage{xcolor}

\usepackage{placeins}

\usepackage{multirow}

\begin{document}

\begin{frontmatter}

%\title{Elsevier \LaTeX\ template\tnoteref{mytitlenote}}

\title{\textcolor{black}{Economic versus energetic model predictive control of a cold production plant with thermal energy storage}}

%\tnotetext[mytitlenote]{Fully documented templates are available in the elsarticle package on \href{http://www.ctan.org/tex-archive/macros/latex/contrib/elsarticle}{CTAN}.}

%% Group authors per affiliation:
%%\author{Elsevier\fnref{myfootnote}}
%\author{Manuel G. Satué$^{1,*}$, Manuel R. Arahal$^{1}$, Luis F. Acedo$^{1}$, Manuel G. Ortega$^{1}$}
%\address{$^{1}$Systems Engineering and Automation Department, ETSI, U. Sevilla, Seville, 41092, Spain }
%%\fntext[myfootnote]{Since 1880.}

% or include affiliations in footnotes:
\author[myaddress]{Manuel G. Satu\'e \corref{mycorrespondingauthor}}
\ead{mgarrido16@us.es}
\author[myaddress]{Manuel R. Arahal}
\author[myaddress]{Luis F. Acedo}
\author[myaddress]{Manuel G. Ortega}
\cortext[mycorrespondingauthor]{Corresponding author}
\address[myaddress]{Systems Engineering and Automation Department, ETSI, U. Sevilla, Seville, 41092, Spain}

%\address[mysecondaryaddress]{360 Park Avenue South, New York}

\begin{abstract}
\textcolor{black}{Economic model predictive control has been proposed as a means for solving the unit loading and unit allocation problem in multi-chiller cooling plants. The adjective economic stems from the use of financial cost due to electricity consumption in a time horizon, such is the loss function minimized at each sampling period. The energetic approach is rarely encountered. This article presents for the first time a comparison between the energetic optimization objective and the economic one. The comparison is made on a cooling plant using air-cooled water chillers and a cold storage system. Models developed have been integrated into \textit{Simscape}, and  non-convex mixed optimization methods used to achieve optimal control trajectories for both energetic and economic goals considered separately. The results over several scenarios, and in different seasons, support the consideration of the energetic approach despite the current prevalence of the economic one. The results are dependent on the electric season and the available tariffs. In particular, for the high electric season and considering a representative tariff, the results show that an increment of about 2.15\% in energy consumption takes place when using the economic approach instead of the energetic one. On the other hand, a reduction in cost of 2.94\% is achieved.}
\end{abstract}

\begin{keyword}
Cooling plant \sep Optimization \sep Predictive control \sep Thermal Storage \sep Scheduling
%\texttt{elsarticle.cls}\sep \LaTeX\sep Elsevier \sep template
%\MSC[2010] 00-01\sep  99-00
\end{keyword}

\end{frontmatter}

%\linenumbers

%%%%%%%%%%%%%%%%%%%%%%%%%%%%%%%%%%%%%%%%%%%%%%%%%%%%%%%%%%%%%%%%%%%%

\section*{Nomenclature}

\begin{table}[h!]
	\begin{center}
		\begin{tabular}{cl}
			Abbreviation & Meaning \\
			\hline
			$EER$ & Energy Efficiency Ratio \\	
			$COP$ & Coefficient Of Performance \\
			$CTES$ & Cold Thermal Energy Storage \\			
			$MPC$ & Model Predictive Control \\
			$OCL$ & Optimal Chiller Loading \\				
			$OCS$ & Optimal Chiller Sequencing \\				
			$PLR$ & Partial Load Ratio \\					
			$TES$ & Thermal Energy Storage \\
		\end{tabular}
	\end{center}
\end{table}

\begin{table}[h!]
	\begin{center}
	\begin{tabular}{cll}
	Variable & Meaning & Units \\
	\hline
	$C_p$ & Specific heat of water & $\mathrm{W/kg \; K}$ \\
	$J$ & Cost of objective function &  \\
	$J^{\star}$ & Augmented cost of objective function &  \\	
	$\dot{m}$ & Mass flow rate &  $\mathrm{kg/s}$ \\
    $M$ & Working mode of the TES system& \\
	$\dot{Q}$ & Thermal power &  $\mathrm{W}$ \\
	$P$ & Electrical power &  $\mathrm{W}$ \\
	$S$ & On/off state & \\
	$T$ & Temperature &  $\mathrm{K}$ \\
	$Tar$ & Electric tariff & \euro/kWh \\
	$\bf{x}$ & Vector of decision variables & \\
	
	\end{tabular}
	\end{center}
\end{table}

\vspace{0.7cm}

\begin{table}[h!]
	\begin{center}
	\begin{tabular}{cl}
	Subscripts & Meaning \\
	\hline
	$chs$ & Chillers\\		
	$env$ & Environment (ambient)\\
	$i$ & Chiller number\\
	$L$ & Load\\
	$T$ & TES\\	
	\end{tabular}
	\end{center}
\end{table}

\vspace{0.7cm}

\begin{table}[h!]
	\begin{center}
	\begin{tabular}{cl}
	Superscripts & Meaning \\
	\hline
	$*$ & Forecasted \\	
	$c$ & Condenser \\
	$e$ & Evaporator \\
	$econ$ & Economic\\	
	$ener$ & Energetic\\
	$i$ & Input \\
	$l$ & Lower\\
	$nom$ & Nominal \\
	$o$ & Output \\
	$u$ & Upper\\
	\end{tabular}
	\end{center}
\end{table}

\vspace{0.7cm}

\begin{table}[h!]
	\begin{center}
\begin{tabular}{cl}
	Quantities & Meaning \\
	\hline
	$n$ & Number of chiller units \\
	$N_p$ & Prediction horizon \\
\end{tabular}
	\end{center}
\end{table}

\clearpage
%%%%%%%%%%%%%%%%%%%%%%%%%%%%%%%%%%%%%%%%%%%%%%%%%%%%%%%%%%%%%%%%%%%%%%%%%%%%%%%
\section{Introduction}
\textcolor{black}{
Many facilities such as markets, food processing plants, hospitals, shopping centers and offices require cooling. Currently, the electricity consumption due to the cold demand represents 10\% of world consumption and 20\%-40\% of total consumption of buildings according to the International Energy Agency \cite{IEA2018}. The demanded and installed refrigeration capacity will continue to grow as countries with emerging economy progress. However, this growing energy demand will have a major impact in the greenhouse effect and global warming. Therefore, it is of the great importance to improve the performance of large production installation.} 

\textcolor{black}{
Cooling plants use several chillers in parallel connection, since this scheme provides flexibility, stand-by capacity, better maintenance and better performance under partial-load conditions \cite{AKBARIDIBAVAR2022111571}. In such circumstances, it is convenient to partition the load among the units in the multi-chiller plant to reduce the electricity consumption. The efficiency, expressed as the coefficient of performance (COP) or energy efficiency ratio (EER), depend on the operating point, and thus must be managed carefully. The classical approach to manage multi-chillers loading is Optimal Chiller Loading (OCL), aiming at solving the thermal load partition  reducing costs the associated costs. Cooling demand forecasts are used to develop loading plans with some lead time \cite{zhou2021incorporating,arahal2021chiller}. Also, different electricity prices during the day can be considered to further reduce costs. Examples of these methods can be found in \cite{saeedi2019robust, shao2019chiller,jabari2020energy}.}

\textcolor{black}{
Thermal Energy Storage (TES) has been proposed to cope with time-varying electricity tariffs. This allows for a reduction in peak electric demand, as the units work part-load with better COP \cite{campos2021optimal}, etc. The OCL problem becomes more difficult as dynamic effects appear. Then, the state at one given time instant affects the loading problem for the following time instants. The Optimal Chiller Sequencing (OCS) problem appears as a result, where the state of each chiller (on/off) is also considered. The  scheduling of chiller units together with the TES operation constitutes an optimal control problem, as it uses a cost function that is a function of the state and control variables \cite{erdemir2021experimental}.}

\textcolor{black}{
The Model Predictive Control (MPC) approach can address the scheduling problem with TES. There, a model and a sliding time window provide a future framework where control actions are chosen to optimize a functional. For instance, a planning algorithm for a chiller based on a predictive model to minimize the electrical cost one day in advance on the market is presented in \cite{balint2018cost}. Notice that the receding horizon strategy does not provide in general a global optimal solution \cite{camacho2013model}, nevertheless the MPC can handle perturbations and model mismatches whereas optimal control can't.}

\textcolor{black}{
In the literature, Exergo-economic analysis has been proposed to identify near-optimal design and operation strategies, simultaneously accounting for capital and operating costs \cite{CATRINI2020113051}. Regarding operation strategies, the so called economic MPC \cite{rawlings2018economic,risbeck2019economic} has been used to seek optimal policies for different systems. In most cases the chillers are modelled using data-sheets from manufacturers, in other cases this data is enhanced with simulations using tools such as TRNSYS \cite{wang2018practical}, and data-driven approaches \cite{ho2021variable}. The impact of different factors, including MPC tuning, has been the subject of several studies. For instance, in \cite{DETTORRE2019524}, the prediction horizon is considered in a heating system using TES. In \cite{WANG2022117809} nonlinear models are used to improve the energy efficiency of a cooling plant. A distributed TES is considered in \cite{taylor2021model} for heating and cooling networks using MPC. Some MCP variants are compared in \cite{li2021comparative} using real meteorological data. The interaction of MPC control of a TES with a smart grid is presented in \cite{tang2019mpc}. In these and most other works, the objective is the economic maximization of benefits and reduction of operating costs. Few publications regarding MPC deal with energy reduction per se, as in \cite{chen2021model} where a comparison with classic control schemes is presented.This work compares such approach with the energetic optimization, i.e. the reduction of the total energy required to supply the cooling demand. Thus, this paper focuses on the comparison of both economic optimization and a energetic optimization. The scenario is that  of a cold production plant for an installation that requires a high thermal power, when a TES system is used.}

\subsection{Novelty and contributions}
\textcolor{black}{
Regarding the previous state of the art, this paper contains a number of novelties and contributions both in the methodology and in the final objective. 
\begin{itemize}
    \item In most of the above cited papers, the power load of each chiller or Part Load Ratio (PLR) is used as an independent variable for the optimization. However, the PLR is not directly controlled but resulting from the combination of manipulated variables such as mass flows and temperature set-points. In this work, the chillers' output temperature references and their mass flow rates are used as independent variables similarly to what is done in  \cite{chiam2019hierarchical,karami2018particle,arahal2021optimal}, but also incorporating the on/off state of chiller units and TES system simultaneously. In this way both the OCL and OCS problems are tackled at once.
    \item Two MPC strategies with two different optimization policies have been developed for the operation of a cold production plant with thermal energy storage considering that low level controllers make the intermediate variables attain their references.
    \item Real data from a Laboratory building is used, including real temperature and real forecasts made by an external governmental agency. In this way, the discrepancies between actual trajectories and forecasts are also real.
    \item Models developed use real data from the chillers' manufacturer integrated into \textit{Simscape} for simulations.
    \item Real electricity tariff for several electric seasons have been used for the simulations. 
    \item A non-convex mixed optimization problem with continuous and binary variables is proposed. The  optimization is solved using genetic algorithms based on crossover operator BLX-$\alpha$ and tournament selection. A custom made real coded genetic algorithm has been modified to incorporate both continuous and binary variables. 
    \item As a result of all the above, the comparison between Economic-MPC and Energetic-MPC is provided. The results over several scenarios, and in different seasons, support the consideration of the energetic approach despite the current prevalence of the economic one. The results are dependent on the electric season and the available tariffs. In particular, for the high electric season and considering a representative tariff, the results show that an increment of about 2.15\% in energy consumption takes place when using the economic approach instead of the energetic one. On the other hand, a reduction in cost of 2.94\% is achieved.
\end{itemize}}

The remainder of the paper is organized as follows: \mbox{Section \ref{Sec_2.Sistema}} presents the cha\-rac\-te\-ris\-tics of the cold production plant, including the  cold water system ge\-ne\-ra\-tion, thermal storage system, and the thermal load. \mbox{Section \ref{Sec_3.Modelado}} presents the modelling of each component of the system. \textcolor{black}{Next, in \mbox{Section \ref{Sec_5.Optimizacion}} the optimization problem is presented.} Simulation results are presented in Section \ref{Sec_6.Resultados}, and finally, the main conclusions  are presented in \mbox{Section \ref{Sec_7.Conclusiones}}.

%%%%%%%%%%%%%%%%%%%%%%%%%%%%%%%%%%%%%%%%%%%%%%%%%%%%%%%%%%%%%%%%%%%%
\section{Cold production plant} \label{Sec_2.Sistema}

\begin{figure}[htbp]
	\centering
	\includegraphics[scale = 0.5]{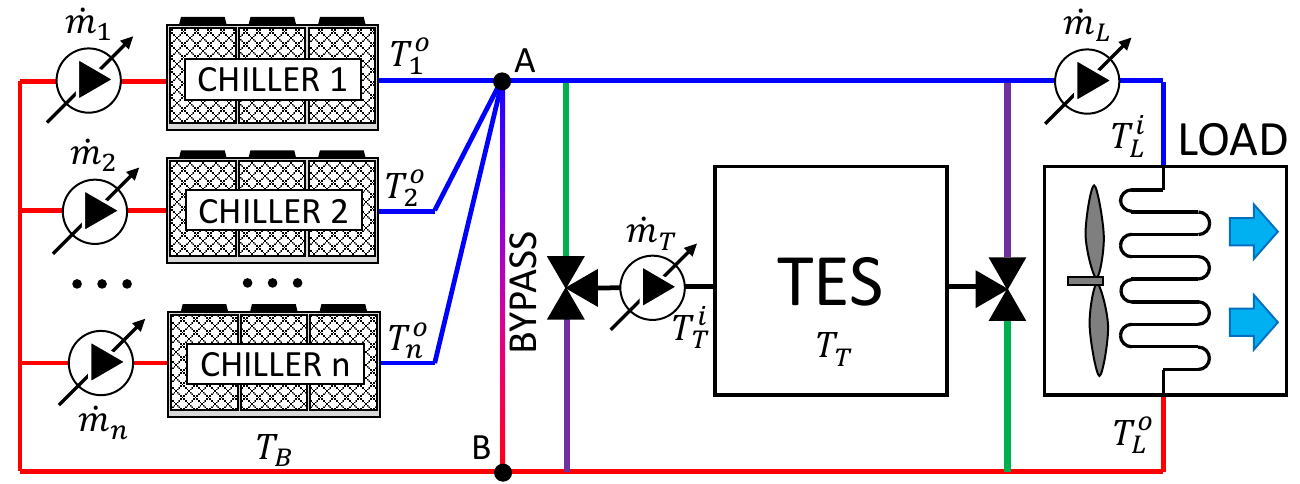}
	\caption{Diagram of the cold generation plant. The two possible configurable  water circuits using the two three-way valves, which are related to the working modes of the TES, are represented in green color for the charging of the TES and in purple color for the discharging case.}
	\label{fig:DiagramaPlanta}	
\end{figure}

The system used in this work (see Figure \ref{fig:DiagramaPlanta}) consists of three main elements: a cold production system, a TES system and a thermal load that is supplied by the cooling plant and TES.

\textcolor{black}{
The cold production system consists of four chiller units corresponding to the RTAC 400, RTAC 300, RTAC 250, and RTAA 125 models from TRANE.} These are air-cooled water chillers. The chosen models have an approximate rated power of 1400, 1060 834 and 365 kilowatt thermal of refrigeration at full load respectively. In these conditions the RTAC chillers have a COP value of 2.8 (including the electrical power of com\-pres\-sors, fans and control systems), and the RTAA chiller has a COP value of 3.1. Working at a partial load of 25\%, the COP values of the RTAC chillers increase to 5.8, 5.33 and 6.06 respectively, while the RTAA model reaches a value of 4.48. More operational data of the chillers working at different conditions and load are shown in Tables \ref{tab:OperationalDataFullLoad} and \ref{tab:OperationalDataPartialLoad}. The low-level control of these machines allows to maintain the outlet temperature of chilled water in the specified reference, being [5, 9] $\mathrm{^o}$C the temperature range in nominal operation. The water flow through each machine is regulated by external pumps of variable flow-rate.

%xxxen rojo 

\begin{table}[t]
{\color{black}
	\caption{Operational data of chillers at full load. ELWT is the \textit{evaporator leaving water temperature} and CAET is the \textit{condenser air entering temperature}.}
	\begin{center}
		\begin{tabular}{c c c c c c}
		    \multicolumn{2}{c}{} & \multicolumn{4}{c}{CAET} \\ \cline{3-6}
		    \multicolumn{2}{c}{}& \multicolumn{2}{c}{30 $\mathrm{^{\circ}C}$} & \multicolumn{2}{c}{45 $\mathrm{^{\circ}C}$} \\ \cline{3-6}
		    ELWT & Model & kW cooling & COP & kW cooling & COP \\ \hline
		    \multirow{4}{*}{5 $\mathrm{^{\circ}C}$} & RTAC 400 & 1407.1 & 3.1 & 1145.9 & 2 \\
		                       & RTAC 300 & 1062.9 & 3.1 & 865.6 & 2 \\
		                       & RTAC 250 & 836.1 & 3.1 & 678.6 & 2 \\
		                       & RTAA 125 & 375.15 & 3.42 & 306.94 & 2.22 \\ \hline
            \multirow{4}{*}{9 $\mathrm{^{\circ}C}$} & RTAC 400 & 1580.1 & 3.2 & 1196.1 & 2.2 \\
		                       & RTAC 300 & 1192.6 & 3.2 & 903.3 & 2.2 \\
		                       & RTAC 250 & 939.1  & 3.2 & 718 & 2.2 \\
		                       & RTAA 125 & 413.48 & 3.6 & 336.83 & 2.37 \\ \hline		                       
		\end{tabular}
		\label{tab:OperationalDataFullLoad}
	\end{center}
	}
\end{table}

\begin{table}[t]
{\color{black}
	\caption{Operational data of chillers at partial load.}
	\begin{center}
		\begin{tabular}{c c c c c}
		    \multicolumn{1}{c}{} & \multicolumn{4}{c}{COP} \\ \cline{2-5}
		    \% Load & RTAC 400 & RTAC 300 & RTAC 250 & RTAA 125 \\ \hline
            100 & 2.75 & 2.78 & 2.75 & 3.07 \\
            75 & 3.72 & 3.72 & 3.69 & 3.54 \\
            50 & 4.42 & 4.04 & 4.68 & 4.33 \\
            25 & 5.82 & 5.33 & 6.06 & 4.48 \\ \hline
		\end{tabular}
		\label{tab:OperationalDataPartialLoad}
	\end{center}
	}
\end{table}

\textcolor{black}{
The energy storage system consists of a water tank with a constant volume of 1000 $\mathrm{m^3}$. Two three-way valves are used to change the operation mode of the TES between charging and discharging,  as illustrated in the diagram of Figure \ref{fig:DiagramaPlanta}.} In the \textit{discharge} case, the cold water stored in the TES is sent to the chilled water line inlet of the building, while the cool water coming from the outlet of the building is pumped into the tank. In the \textit{charge} case, chilled water is pumped to the tank while the not too cool water inside the tank is sent to mix with the outlet cool water of the building.

\textcolor{black}{
The thermal load comes from laboratories buildings of the \textit{Escuela Técnica Superior de Ingenieria} in Seville, Spain.} This load varies in time depending on the occupancy level of the buildings, weather conditions outside, etc. The function of the \textit{bypass} \mbox{(see Figure \ref{fig:DiagramaPlanta})} is to adjust the range of flows that the facility may demand to the range of flows in which chillers can operate.

\FloatBarrier

%%%%%%%%%%%%%%%%%%%%%%%%%%%%%%%%%%%%%%%%%%%%%%%%%%%%%%%%%%%%%%%%%%%%
% \section{Modelling} \label{Sec_3.Modelado}

\subsection{Modelling} \label{Sec_3.Modelado}
\begin{figure}[htbp]
	\centering
	\includegraphics[scale = 0.5]{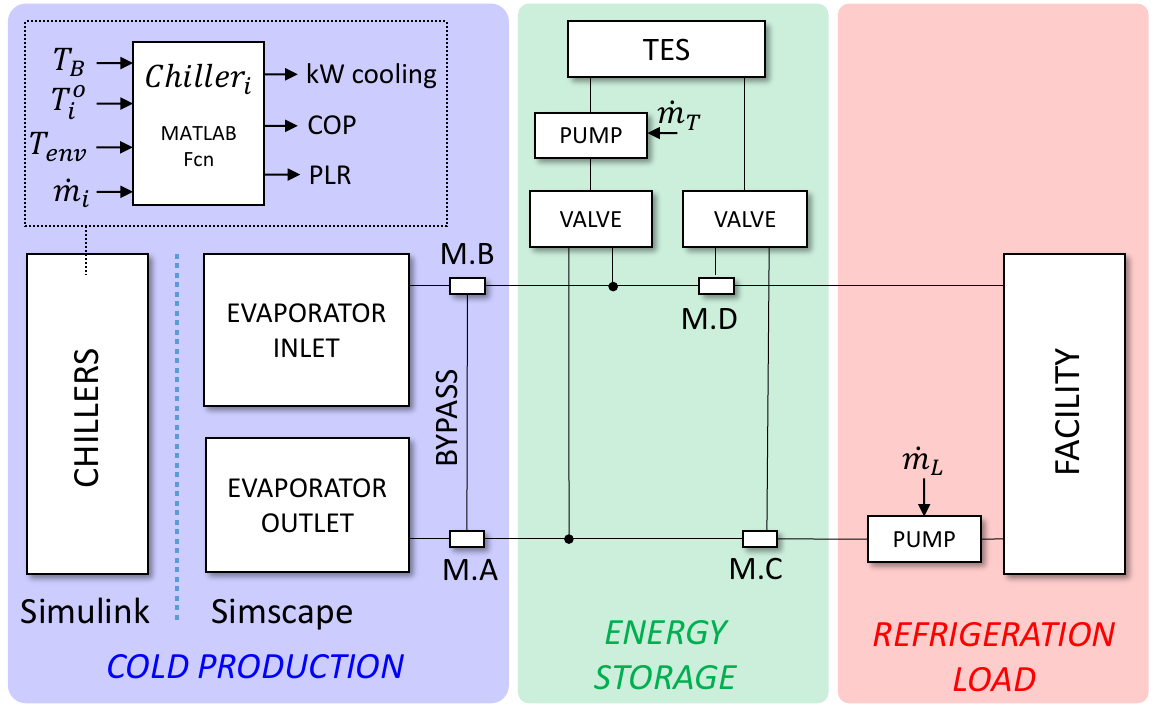}
	\caption{Block diagram of the system in Simscape.}
	\label{fig:DiagramaBloquesSimscape}	
\end{figure}

The modelling of the cold production plant has been developed in the \mbox{\textit{Simscape}} simulation environment of \textit{MATLAB's Simulink} \textcolor{black}{(MATLAB 9.10, Simulink 10.3, Simscape 5.1 (2021a))}. \textcolor{black}{Figure \ref{fig:DiagramaBloquesSimscape} depicts the block diagram of the model. In this figure the three main parts of the model are highlighted in blue color (cold production), green color (energy storage) and red color (refrigeration load). The principal components that are part of the production plant have been developed as submodels which are explained in the following subsections.}

\FloatBarrier

\subsection{Chillers subsystem}

Manufacturers provide cooling thermal power and COP for each individual machine model, both at full and partial load for various combinations of water temperature at the outlet of the evaporator and air temperature at the inlet of the condenser. In this work, the modelling of each chiller is based on operating data provided by the manufacturer. The model interpolates COP values depending on water temperature at the evaporator outlet, air temperature at the condenser inlet and the load ratio of the machine (PLR, \textit{part load ratio}). The electrical power consumed by the machine is calculated with the obtained values of cooling thermal power and COP.

The cooling power provided by the i-th chiller, $\dot{Q}_{i}$, can be expressed as
\begin{equation}
	\label{eq:Chiller_PotenciaTermica}
	\dot{Q}_{i} = \dot{m}_{i}{C}_{p}({T}^{i}_{i}-{T}^{o}_{i}),
\end{equation}
where $\dot{m}_{i}$ is the mass flow rate, $C_p$ is the specific heat of water, ${T}^{i}_{i}$ is the temperature of water at the evaporator inlet of the chiller and ${T}^{o}_{i}$ is the temperature at the evaporator outlet. The part load ratio of a chiller is defined as

\begin{equation}
	\label{eq:Chiller_PLR}
	{PLR}_{i} = \frac{\dot{Q}_{i}}{\dot{Q}^{nom}_{i}}  
\end{equation}

The coefficient of performance is a function of several variables which is obtained by interpolating data provided by the manufacturers defined as

\begin{equation}
	\label{eq:Chiller_COP}
	{COP}_{i} = f(PLR_i, T^{e}_{i},T^{c}_{i}),
\end{equation}

where $T^{e}_{i}$ is the temperature of water at the evaporator outlet and $T^{c}_{i}$ is the tempertature of air at the condenser inlet.

The electric power consumed by the chiller, $P_i$, is computed used the definition of COP:

\begin{equation}
	\label{eq:Chiller_PotenciaElectrica}
	{P}_{i} = \frac{\dot{Q}_{i}}{{COP}_{i}}
\end{equation}

The provided COP includes the anciliary power consumed by the chiller, such as control electronics consumption.

Finally, the temperature of chilled mixed water leaving the chillers group is obtained by the following expression

\begin{equation}
	\label{eq:Chiller_Tmix}
	{T}^{o}_{M} = \frac{\sum_{i=1}^{n}{\dot{m}_i {T}^{o}_{i}}}{\sum_{i=1}^{n}{\dot{m}_i }}
\end{equation}

As can be seen in Figure \ref{fig:DiagramaBloquesSimscape}, the model of the set of chillers is programmed in a \textit{MATLAB Function} block. The model inputs of a chiller are: water tem\-pe\-ra\-tu\-re at point B of \textit{bypass}, reference temperature of water at the evaporator outlet, water mass flow through the evaporator and environment tem\-pe\-ra\-tu\-re. As outputs it provides: COP, electrical power consumed and cooling thermal power. The \mbox{parallel} connection of the chillers is also done with a \textit{MATLAB Function} block that receives as inputs the evaporator outlet temperature of each machine together with its corresponding mass flow rates, and returns as output variables the total mass flow rate of the set of chillers and the mixing temperature of cooled water. The subsystem of chillers also includes the \textit{bypass} required by the system to comply with the mass conservation balance. In the Simscape simulation environment, the bypass consists of two blocks, M.A and M.B in Figure \ref{fig:DiagramaBloquesSimscape}, in which separate mass and energy balances are established between the respective input/output ports.

\subsection{Energy storage subsystem}

The energy storage subsystem is based on the accumulation of mass and energy in a chamber containing a fixed volume of water. The tank presents an inlet and outlet port, through which water flows. The mass of the fluid varies with density, a property that is generally a function of pressure and temperature. 

The flow resistance between the inlet and the interior of the chamber is assumed to be negligible. Therefore, the pressure in the interior is equal to the pressure at the inlet.

To simulate the energy storage subsystem it is also necessary to model, in addition to the elements for the charge and discharge of the TES, a pump and a three-way valve.

\textbf{Mass balance}

The volume of the chamber is fixed, but the compressibility of the fluid means that its mass can change with pressure and temperature. The rate of mass accumulation in the chamber must equal the mass flow rates in through inlet and outlet ports as stated in Equation (\ref{eq:BalanceMasa}),

\begin{equation}
	\label{eq:BalanceMasa}
	\left( \frac{1}{\beta} \frac{dp}{dt} - \alpha \frac{dT}{dt} \right) \rho V = \dot{m}^{i} + \dot{m}^{o}
\end{equation}

where the left-hand side is the rate of energy accumulation and $p$ is the pressure $\mathrm{[MPa]}$, $T$ is the temperature $\mathrm{[K]}$, $\beta$ is the isothermal bulk modulus $\mathrm{[MPa]}$, $\alpha$ is the isobaric thermal expansion coefficient $\mathrm{[1/K]}$ and $\dot{m}$ is the mass flow rate $\mathrm{[kg/s]}$.\\

\textbf{Energy balance}

The rate of energy accumulation in the internal fluid volume must therefore equal the sum of the energy flow rate in through inlet and outlet ports as stated in Equation (\ref{eq:BalancEnergia}),

\begin{equation}
	\label{eq:BalancEnergia}
	\left[ \left( \frac{h}{\beta} - \frac{T\alpha}{\rho} \right) \frac{dp}{dt} + \left( C_p - h\alpha \right) \frac{dT}{dt}  \right] \rho V = \phi^{i} + \phi^{o}
\end{equation}

where the left-hand side is the rate of energy accumulation and $h$ is the enthalpy $\mathrm{[kJ/kg]}$, $\rho$ is the density $\mathrm{[kg/m^{3}]}$, $C_p$ is the specific heat $\mathrm{[kJ/kg\;K]}$, $V$ is the chamber volume $\mathrm{[m^3]}$ and $\phi$ is the energy flow rate $\mathrm{[kW]}$.\\

\subsection{Facility subsystem}

The modelling of the \textcolor{black}{refrigeration load (highlighted in red color in Figure \ref{fig:DiagramaBloquesSimscape})} block has been simulated by a black-box type model. The load block consists of a pipe through which water circulates and where it is allowed a heat exchange with the environment through a thermal port. The load flow rate, $\dot{m}_L$, represents the inputs variable of the block. The load profiles of the installation are variable with the time, and through a physical signal are connected with the block.

%%%%%%%%%%%%%%%%%%%%%%%%%%%%%%%%%%%%%%%%%%%%%%%%%%%%%%%%%%%%%%%%%%%%
\section{\textcolor{black}{Control and Optimization}} \label{Sec_5.Optimizacion}

\textcolor{black}{The cooling plant under MPC control relies on low-level controllers that accept reference values as inputs and produce the adequate control actions in real-time. These control actions are the command signals for pumps to produce the required mass flows and the command signals for chillers to produce the required output water temperature. The low level control is relatively fast with  closed-loop dynamics of the order of a few tens of seconds, whereas other processes related to optimization take hours. The simulations take advantage of this fact by replacing fast dynamics by static terms. This do not affect the optimization result and allow speeding up the simulations.}

\textcolor{black}{The MPC is regarded as a high-level layer in charge of sending references to low-level controllers that drive the pumps of the chillers, the TES, and building circuit.} It uses a model of the plant, which is simulated in a sliding time window associated to the prediction horizon, $N_p$, \textcolor{black}{to minimize a figure associated to the merit of the solution. In the case of optimizing with a economic criteria, this figure is the economic cost (in monetary units) of the electrical consumption of the chillers in the current time window. In the case of optimizing with an energetic criteria, the figure is the electrical energy cost (in energy units) related to the consumption of the chillers. As in the Model Predictive Control strategy, only the first control action of the computed future control actions' trajectory will be applied}.

\textcolor{black}{The optimization problem linked to these MPC formulations is to decide the operation point for:  each chiller, the TES, and the load. This has to be done  for each time period into which the prediction window is divided, in this case 24 hours with one hour intervals.}

The operation point for a single period is defined by the temperature references at the evaporator outlet of each chiller, $T^{o}_{i}$, the mass flow rates through the chillers, $\dot{m}_i$, binary variables indicating the current state of the chiller units (on or off), $S_i$, the mass flow rate through the load, $\dot{m}_L$, and through the TES, $\dot{m}_{T}$, and two more binary variables that indicate the on/off state of the TES and its working mode (charging or discharging), $S_T$ and $M$ respectively.

The operation point for all periods of the prediction window is the vector of independent variables $\bf{x}$ defined in Equation (\ref{eq:VarIndep}).
\begin{equation}
	\label{eq:VarIndep}
	\begin{split}
		{\bf x} =
		\bigl[ 
		\dot{m}_1{|}^{Np}_{1},...,\dot{m}_4{|}^{Np}_{1},\\T^{o}_1{|}^{Np}_{1},...,T^{o}_{4}{|}^{Np}_{1},\\S_1{|}^{Np}_{1},...,S_4{|}^{Np}_{1},\\ \dot{m}_L{|}^{Np}_{1},\dot{m}_{T}{|}^{Np}_{1},{S}_{T}{|}^{Np}_{1},M{|}^{Np}_{1}
		\bigr]
	\end{split}
\end{equation}
Please notice that in definition \ref{eq:VarIndep}, the symbol ${|}^{Np}_{1}$ denotes a vector that represents a set of points within a trajectory. For a generic variable $v$ the following definition can be given
\begin{equation}
	\label{eq:tray_gen}
	v{|}^{N_p}_{1}=[v(1),v(2),...,v(N_p)].
\end{equation}

\textcolor{black}{ The aim of the MPC is to provide sufficient cooling power to meet the cooling demand with the lowest possible value for the objective considered: in the case of economic-MPC the objective is reducing the electricity cost, in the case of energetic-MPC the objective is to reduce the energy consumption. Forecasts of of the cooling power demand, $\dot{Q}^{*}_{L}$, and the environment temperature, $T^{*}_{env}$, over the prediction horizon are neede for the MPC formulations.In the case of economic-MPC, the electricity rate must be considered for each time period as it is variable with the time of day. }

\begin{figure}[htbp]
	\centering
	\includegraphics[scale = 0.4]{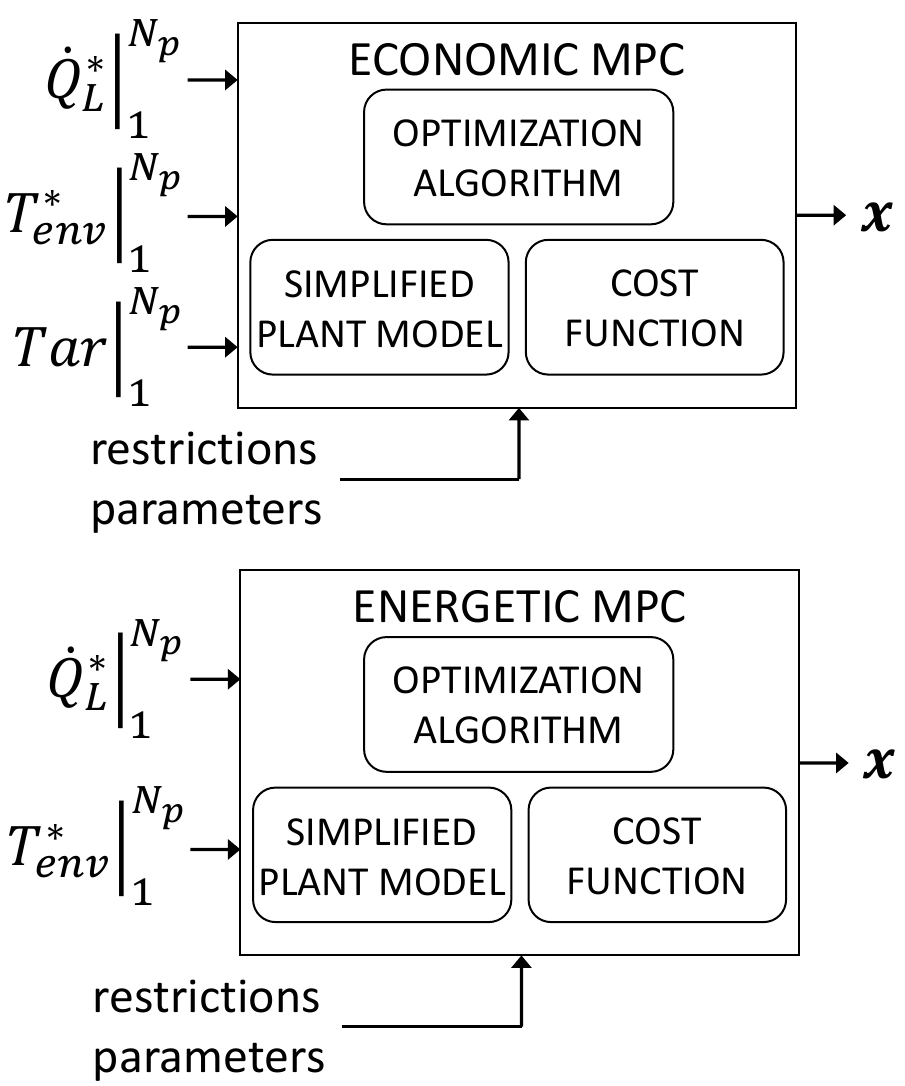}
	\caption{\textcolor{black}{Block diagrams of the Economic and the Energetic optimizers.}}
	\label{fig:DiagramaOptimizador}	
\end{figure}

\subsection{Mathematical Formulation}

\textcolor{black}{As previously stated, both Economic and Energetic MPC try to} minimize a figure or cost which is a function of the chillers' electricity consumption. For the Economic MPC, this cost function is the product of the power consumed by the machines and the corresponding electricity tariff, $Tar$, at each period of the prediction sliding time window according to the Equation
\begin{equation}
	\label{eq:Coste_econ}
	J^{econ} \left({\bf x}\right) = \sum_{k=1}^{Np} \left( \sum_{i=1}^{4} P_{i|k} \cdot Tar_{k} \right).
\end{equation}
For the Energetic MPC case, the cost function is the electrical energy consumed by the chillers in the prediction sliding time window according to the Equation
\begin{equation}
	\label{eq:Coste_ener}
	J^{ener} \left({\bf x}\right) = \sum_{k=1}^{Np} \left( \sum_{i=1}^{4} P_{i|k} \right).
\end{equation}

\textcolor{black}{Please notice that as the prediction window is divided in one hour time intervals, the time interval has been omitted in Equations \ref{eq:Coste_econ} and \ref{eq:Coste_ener}. Furthermore, notice that the electricity tariffs have a resolution of one hour, so the units of both economic and energetic costs functions are coherent.}

\textcolor{black}{
With the given definitions, the optimization problem (either economic or energetic) can be defined mathematically as}
\begin{equation}
	\label{eq:PbOptimizacion}
	\begin{aligned}
		\min_{ {\bf x}} \quad & J( {\bf x}) \\
		\textrm{s.a.} \quad &  x^{l} \le {\bf x}  \le x^{u}\\
		& \dot{Q}( {\bf x})  = \dot{Q}_L  \\
		&  g^{l} \le G({\bf x})  \le g^{u}  \\		
	\end{aligned}
\end{equation}

where the independent variable $\bf x$ defined in (\ref{eq:VarIndep}) is a vector that contains the paths from $k=1$ to $k=Np$ that define the operation point of the system in the prediction horizon.

In the definition of the optimization problem the following \textit{hard} restrictions are used:
\begin{itemize}
	\item $x^{l} \le {\bf x}  \le x^{u}$, to limit independent variables to the physical limits imposed by the plant, such as minimum and maximum flow rates that can circulate by chillers, and the minimum and maximum water temperature at the evaporator outlet, etc. Table \ref{tab:VariableBounds} shows the used decision variables bounds.
	\item $\dot{Q}( {\bf x})  = \dot{Q}_L$, to force demand to be satisfied in each period in which the prediction horizon is divided. This restriction is treated as a \textit{hard} restriction.
	\item $g^{l} \le G({\bf x})  \le g^{u}$, is a set of constraints expressed in terms of functions of ${\bf x}$. $G$ is a vector function which allows to include operation requirements as for example \textcolor{black}{the maximum possible mass flow rate through the refrigeration load once fixed the flow rates through each chiller in a optimization time instant (mass balance).}
	\item At least one machine must be on at any given time.
\end{itemize}

\begin{table}[t]
	\caption{Bounds for Continuous Decision Variables.}
	\begin{center}
		\begin{tabular}{lccc}
			$x$ & $x^l$ & $x^u$ & units\\ \hline
			
			$\dot{m}_1$ & 34 & 105 & kg/s\\
			$\dot{m}_2$ & 20 & 68 & kg/s \\
			$\dot{m}_3$ & 15 & 47 & kg/s \\
			$\dot{m}_4$ & 9.5 &  28.4 & kg/s \\
			$T^{o}_1$ & 5 & 9 & $\mathrm{^{o}C}$\\
			$T^{o}_2$ & 5 & 9 & $\mathrm{^{o}C}$\\
			$T^{o}_3$ & 5 & 9 & $\mathrm{^{o}C}$\\
			$T^{o}_4$ & 5 & 9 & $\mathrm{^{o}C}$\\
			$\dot{m}_L$ & 9.5 & 268.4 & kg/s \\
			$\dot{m}_{T}$ & 1 & 50 & kg/s \\
		\end{tabular}
		\label{tab:VariableBounds}
	\end{center}
\end{table} 

{\color{black}Furthermore, a set of \textit{soft} restrictions $h^{l} \le H({\bf x})  \le h^{u}$ has been considered in the cost function by means of penalization coefficients.
} 

\begin{itemize}
	\item The inlet water temperature of a chiller's evaporator must be higher than its outlet temperature
	\begin{equation}
		\label{eq:r1}
		T^{i}_{i} > T^{o}_{i} \quad \forall \; i
	\end{equation}
	\item The load inlet water temperature must not exceed a prefixed maximum value
	\begin{equation}
		\label{eq:r2}
		T^{i}_{L} < T^{u}_{L}
	\end{equation}
	\textcolor{black}{where $T^{u}_{L}=15\;^{\circ}\mathrm{C}$.}
	\item The temperature of the water inside the TES must not exceed a given maximum value
	\begin{equation}
		\label{eq:r3}
		T_{T} < T^{u}_{T}
	\end{equation}
	\textcolor{black}{where $T^{u}_{T}=15\;^{\circ}\mathrm{C}$.}
	\item The absolute temperature difference between the input and output of the evaporator of each machine must belong to a certain pre-established range
		\begin{equation}
		\label{eq:r4}
		T^{i}_{i}-T^{o}_{i} \in [\Delta{T}^{l},\Delta{T}^{u}] \quad \forall \; i
	\end{equation}
	\textcolor{black}{where $\Delta{T}^{l} = 3.3\;^{\circ}\mathrm{C}$ and $\Delta{T}^{u} = 10\;^{\circ}\mathrm{C}$.}
\end{itemize}

\textcolor{black}{The penalties related to the soft restrictions consist in the addition of an extra cost that increases  when a restriction, $h_i\left({\bf x}\right)$, is violated. Equations (\ref{eq:CosteAumentado}) and (\ref{eq:max_restricc}) define the augmented cost, $J^{*}$, where $\mu_i$ are weight parameters.}
{\color{black}
\begin{equation}
	\label{eq:CosteAumentado}
	J^{\star} \left({\bf x}\right) = J \left({\bf x}\right) + \mu_{i} \cdot \sum_{i \in \tau}h^{+}_{i}\left({\bf x}\right)^2
\end{equation}

\begin{equation}
	\label{eq:max_restricc}
	h^{+}_{i}\left({\bf x}\right)=max\left\{0,h_{i}\left({\bf x}\right)\right\}
\end{equation}
}

\subsubsection{Matlab implementation}
\textcolor{black}{
The optimization problem is characterized by a number of traits that must be considered by the numeric solver. These are: a large number of decision variables, probable existence of multiple local minima, non-availability of the analytical Jacobian and inclusion of binary decision variables. To cope with this, a real-coded genetic algorithm has been selected. The algorithm is custom made and uses the {\em BLX-$\alpha$} crossover operator and tournament-based selection \cite{singh2018modified}. This crossover operator has high capacity of exploration of the solutions space. The values selected for the parameters that configure the genetic algorithm are a population of 3000 individuals, a tournament size of 69 individuals, a mutation ratio equal to 10\% of the population and $\alpha = 0.5$ for the BLX-$\alpha$ operator.}

%%%%%%%%%%%%%%%%%%%%%%%%%%%%%%%%%%%%%%%%%%%%%%%%%%%%%%%%%%%%%%%%%%%%
\section{Simulation results} \label{Sec_6.Resultados}

In order to \textcolor{black}{compare the behaviour of the proposed optimization strategies, a set of  different scenarios will be used. For this kind of facilities with large electric power demand, grid suppliers offer specific electricity rates.} In the region of application, these tariffs establish six consumption time periods with different prices in which  period P1 is the most expensive and  period P6 the cheapest, as shown in Table \ref{tab:ElectricityRates}. The rates also depend on electric seasons (high, medium high, medium and low), which are month-based, and on the day of the week. Also, even though there are six different periods, for any given day only three of them can be applied.

\subsection{\textcolor{black}{Seasons Description}}

\textcolor{black}{The simulations carried out cover three electric seasons (high, medium and low) in which a whole month is simulated for each of them (July, September and May respectively). Table \ref{tab:ElectricSeasons} summarizes this information.}

%The specific electricity tariffs prices used in the simulations are shown in Table \ref{tab:ElectricityRates}.

\begin{table}[t]
	\caption{\textcolor{black}{Electricity rates [\euro/kWh].}}
	\begin{center}
		\begin{tabular}{ccccccc}
			Tariff & P1 & P2 & P3 & P4 & P5 & P6\\ \hline\\[-1.5ex]
			A & 0.2998 & 0.1606 & 0.1368 & 0.1188 & 0.0985 & 0.0991 \\
			B & 0.1802 & 0.1606 & 0.1368 & 0.1188 & 0.0985 & 0.0991 \\
			C & 0.1395 & 0.1278 & 0.1110 & 0.1014 & 0.0927 & 0.0871 \\						
		\end{tabular}
		\label{tab:ElectricityRates}
	\end{center}
\end{table} 

\begin{table}[t]
	\color{black}{
	\caption{Considered electric seasons.}
	\begin{center}
		\begin{tabular}{ccc}
			Month & Season & Periods\\ \hline\\[-1.5ex]
			July & high & P1, P2, P6 \\
			September & medium & P3, P4, P6 \\
			May & low & P4, P5, P6 \\
		\end{tabular}
		\label{tab:ElectricSeasons}
	\end{center}
}
\end{table}

{\color{black}For each of the considered electric seasons, a cooling demand profile for a whole month has been used. These profiles show a dependency on building occupancy level, which is very high on working days during the morning hours, high in the afternoon and very low for the rest of periods. The real cooling demand profile, $\dot{Q}_L$ is approximated by an hourly forecast profile  $\dot{Q}^{*}_{L}$. Uncertainty is considered in the forecast as shown in Figure \ref{fig:TrayectoriasEntradaSimulacion}. Similarly, ambient temperature profiles, $T_L$, have been obtained from real data resources at the Spanish State's Agency of Meteorology \cite{AEMET} database. In addition, hourly temperature forecasts $T^{*}_{L}$, have been  gathered from the AEMET database. 

The following subsections show the results obtained in the simulations of the three electrical seasons. For the high electricity season, a more detailed information is provided as it is the one with more cooling demand.
}

\subsection{High electric season results}

\begin{figure}[h!]
	\centering
	\includegraphics[width = 12cm]{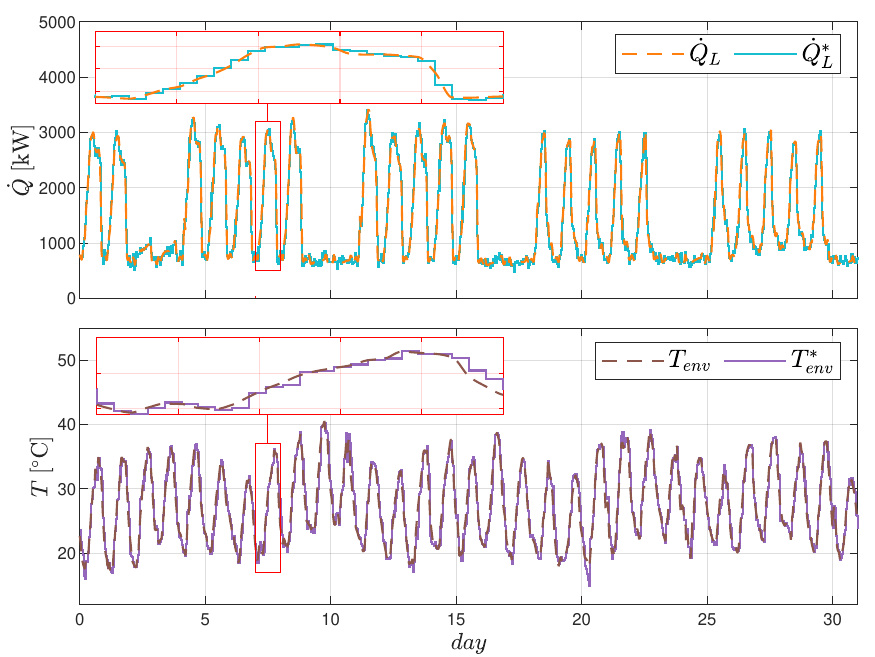}
	\caption{\textcolor{black}{Simulation input profiles. Real and forecasted cooling load ($\dot{Q}_L$ and $\dot{Q}^{*}_{L}$ respectively) and real and forecasted environment temperature ($T_{env}$ and $T^{*}_{env}$ respectively). Detail views for one day are shown in red subgraphs.}}
	\label{fig:TrayectoriasEntradaSimulacion}	
\end{figure}

{\color{black}The input profiles of the simulation for the month of July are shown in Figure \ref{fig:TrayectoriasEntradaSimulacion}. The month of July is a \textit{high electric season}, so the three active time periods are P1, P2 and P6.}

\FloatBarrier

\subsubsection{Energetic optimization}
\begin{figure}[h!]
\begin{subfigure}{1\textwidth}
	\centering
	\includegraphics[width=12cm]{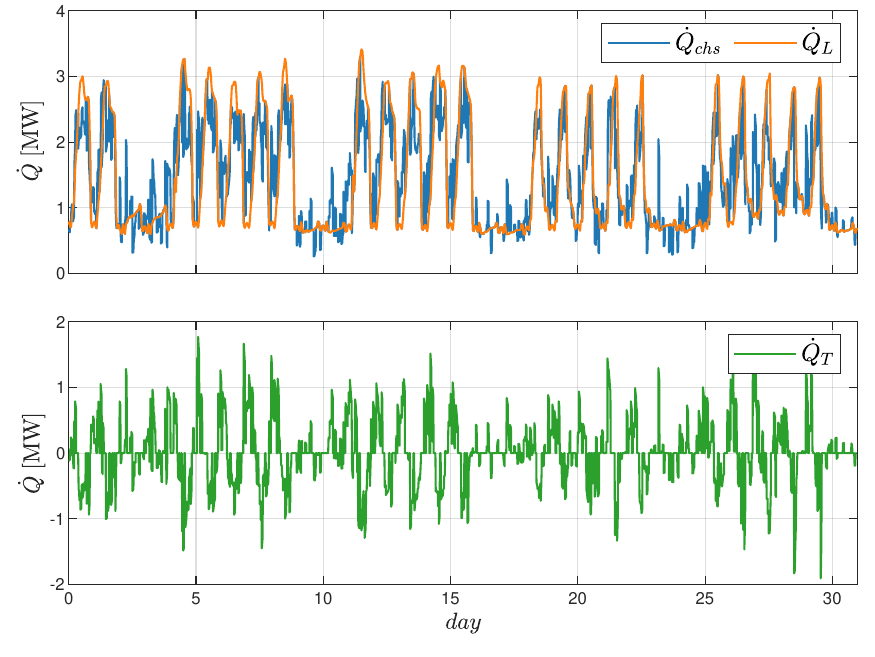}
	\caption{Whole month.}
	\label{fig:BalancedePotencias_ENERGETICO1}
\end{subfigure}
\newline
\begin{subfigure}{1\textwidth}
	\centering
	\includegraphics[width=12cm]{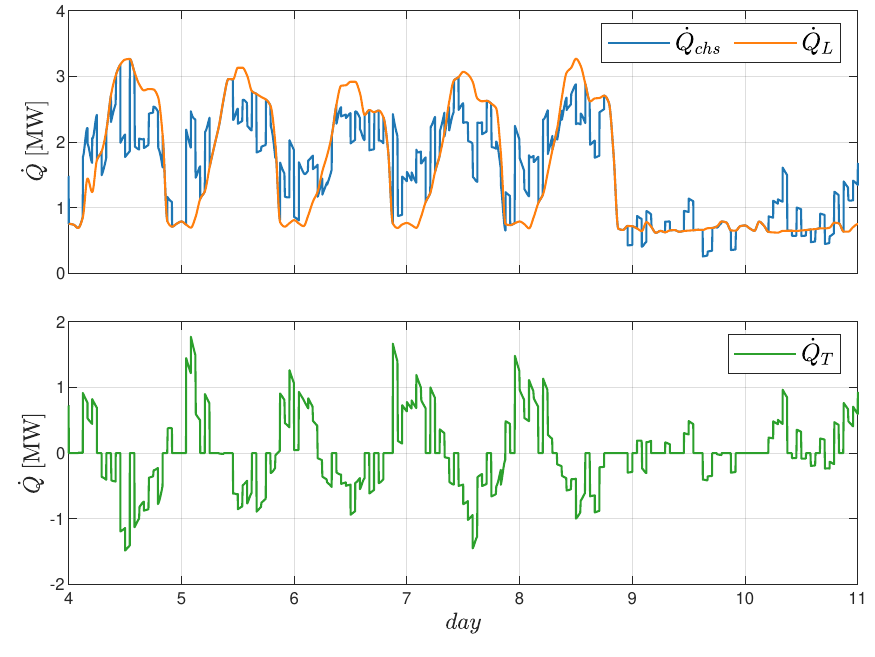}
	\caption{Detail view. A week.}
	\label{fig:BalancedePotencias_ENERGETICO_detalle}
\end{subfigure}
\caption{\textcolor{black}{Cooling power of chillers, load and TES resulting of Energetic MPC optimization.}}
\label{fig:BalancedePotencias_ENERGETICO}
\end{figure}

Figure \ref{fig:BalancedePotencias_ENERGETICO} shows the cooling load demanded by the building, together with the power provided by the chillers, $\dot{Q}_{chs}$, and by the TES, $\dot{Q}_{T}$, when operating the plant with the Energetic MPC. The negative power of the TES means that it is discharging, and positive power that it is charging. It can be clearly seen how the optimizer choose to discharge the TES during the peak thermal load periods relieving the chillers, and to charge it during off-hours.

\begin{figure}[h!]
\begin{subfigure}{1\textwidth}
	\centering
	\includegraphics[width=12cm]{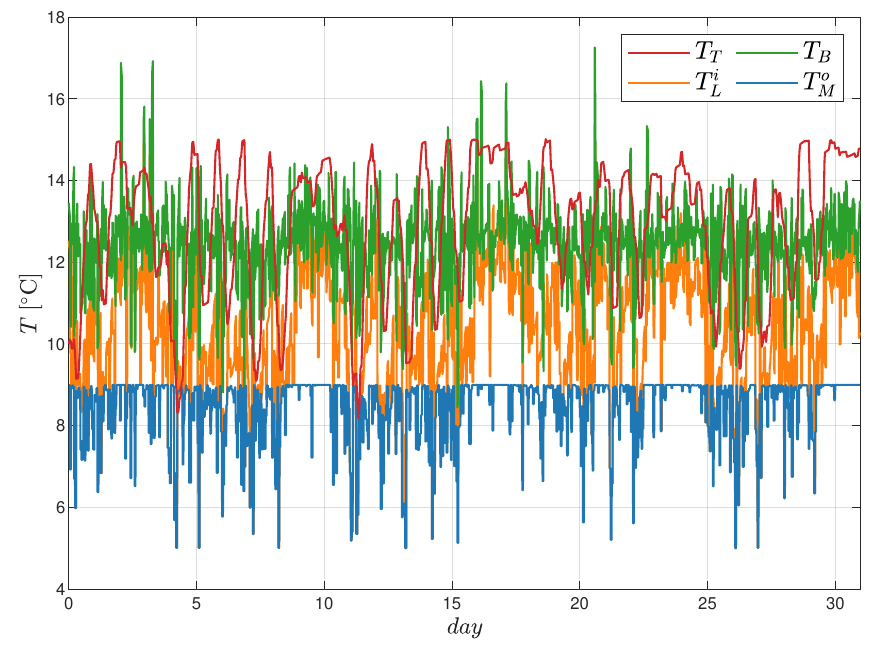}
	\caption{Whole month.}
	\label{fig:Temperaturas_ENERGETICO_A}
\end{subfigure}

\begin{subfigure}{1\textwidth}
	\centering
	\includegraphics[width=12cm]{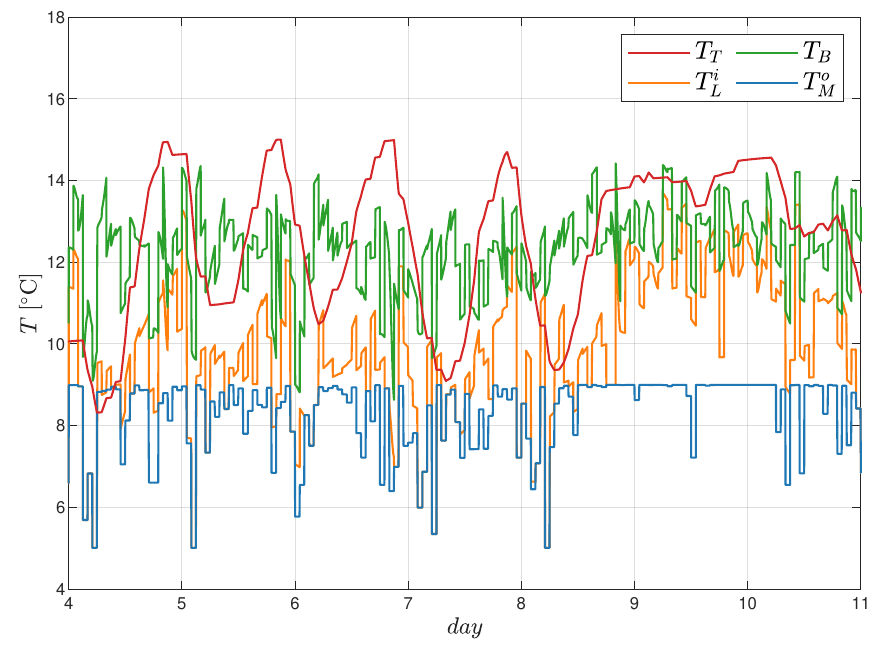}
	\caption{Detail view. A week.}
	\label{fig:Temperaturas_ENERGETICO_B}
\end{subfigure}
\caption{\textcolor{black}{Temperatures resulting of Energetic MPC optimization.}}
\label{fig:Temperaturas_ENERGETICO}
\end{figure}

The temperatures of the water in different points of the cold production plant are shown in Figure \ref{fig:Temperaturas_ENERGETICO}. The action of the restrictions added to the optimization problem can be seen in, for example, the temperature of water inside the TES and the temperature at the input of the load, both always under 15 $\mathrm{^{\circ}C}$.

\FloatBarrier

\subsubsection{Economic optimization}

\begin{figure}[h!]
	\begin{subfigure}{1\textwidth}
		\centering
		\includegraphics[width = 12cm]{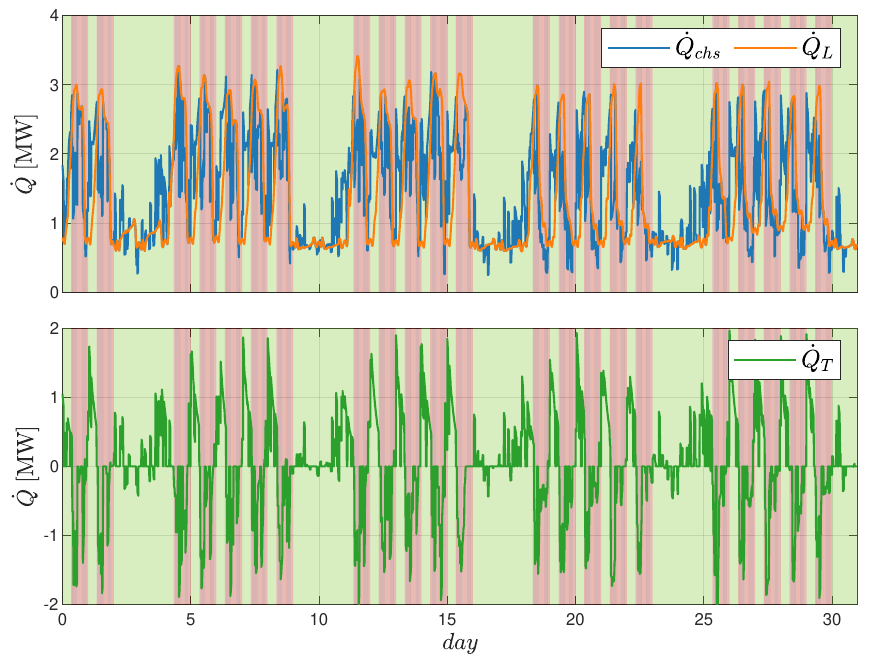}
		\caption{Whole month.}
		\label{fig:BalancedePotencias_ECONOMICO_A}
	\end{subfigure}
	\newline
	\begin{subfigure}{1\textwidth}
		\centering
		\includegraphics[width = 12cm]{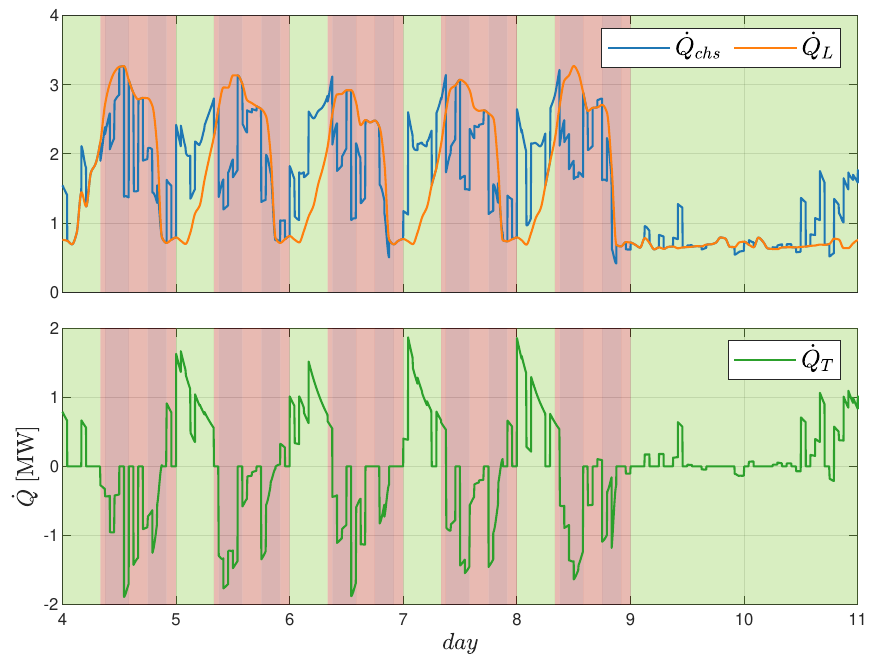}
		\caption{Detail view. A week.}
		\label{fig:BalancedePotencias_ECONOMICO_detalle}	
	\end{subfigure}
	\caption{\textcolor{black}{Cooling power of chillers, load and TES resulting of Economic MPC optimization with electricity tariff A.}}
	\label{fig:BalancedePotencias_ECONOMICO}
\end{figure}

Figure \ref{fig:BalancedePotencias_ECONOMICO} shows the cooling load demanded by the building, together with the power provided by the chillers, $\dot{Q}_{chs}$, and by the TES, $\dot{Q}_{T}$, when operating the plant using the Economic MPC with the electricity tariff A from Table \ref{tab:ElectricityRates}. It can be clearly seen how the optimizer choose to discharge the the TES in the expensive time periods (represented as dark red and red bars) relieving the chillers, and to charge it during the cheap time periods (represented as green bars).

\begin{figure}[h!]
	\begin{subfigure}{1\textwidth}
		\centering
		\includegraphics[width = 12cm]{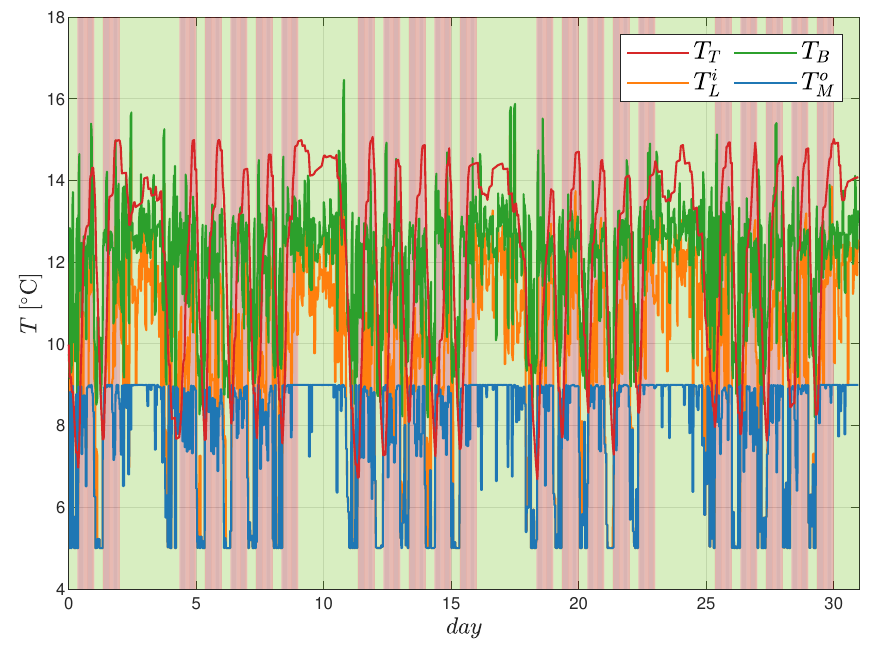}
		\caption{Whole month.}
		\label{fig:Temperaturas_ECONOMICO_A}
	\end{subfigure}
	\newline
	\begin{subfigure}{1\textwidth}
		\centering
		\includegraphics[width = 12cm]{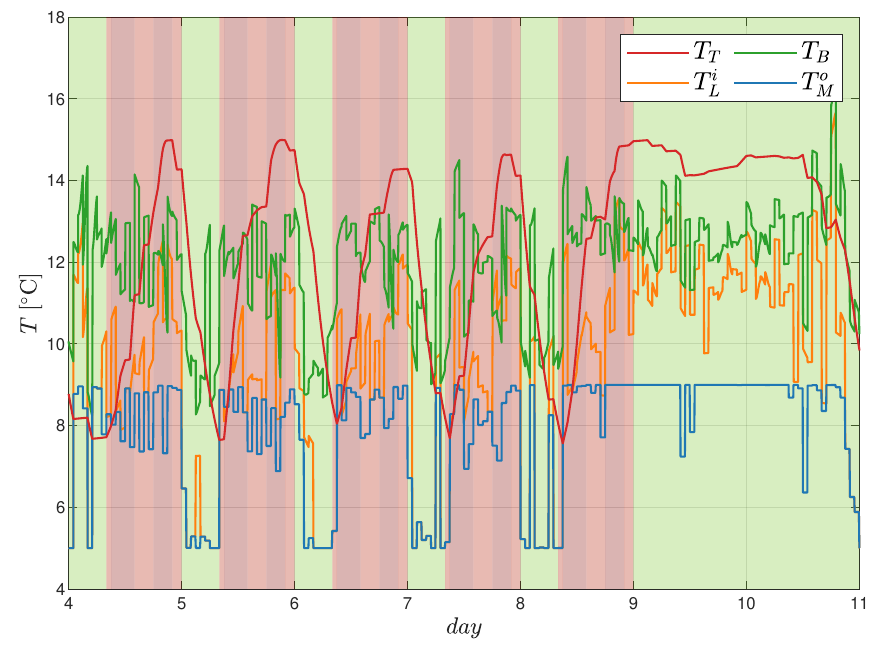}
		\caption{Detail view. A week.}
		\label{fig:Temperaturas_ECONOMICO_detalle}	
	\end{subfigure}
	\caption{\textcolor{black}{Temperatures resulting of Economic MPC optimization with electricity tariff A.}}
	\label{fig:Temperaturas_ECONOMICO}
\end{figure}

Figure \ref{fig:Temperaturas_ECONOMICO} shows a cyclic behaviour of the water temperature inside the TES, $T_{T}$, as occurs in the Energetic optimization case. \textcolor{black}{This temperature grows in the expensive price periods and decays in the cheap periods, as result of charge/discharge cycles of the TES. Please notice how on weekends, when the tariff is the cheapest and the cooling load is very low, the optimizer decides not to fully charge the TES until the last moment.}

\FloatBarrier

\subsubsection{Energetic vs Economic comparison}

\begin{figure}[h!]
	\centering
	\includegraphics[width=12cm]{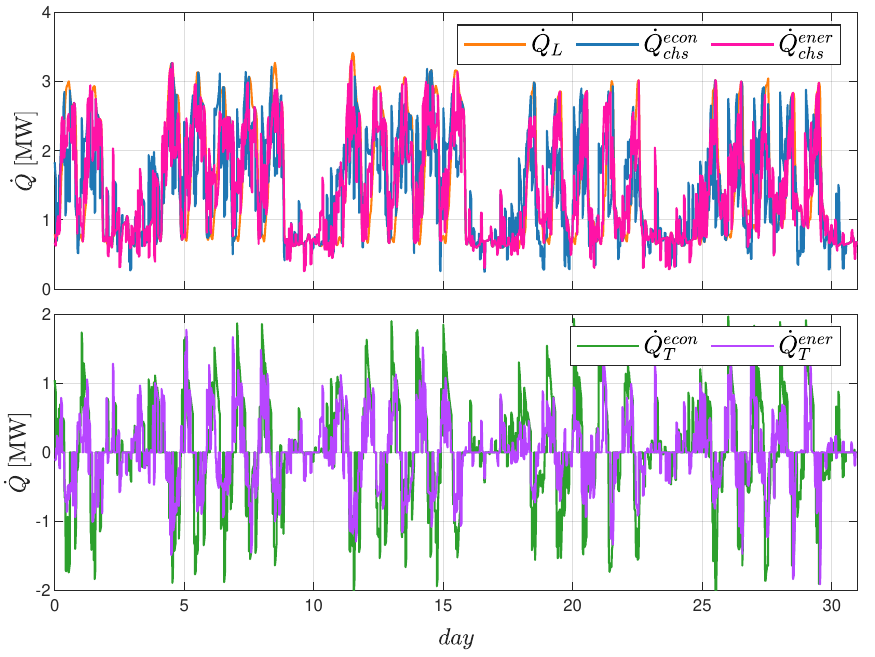}
	\caption{\textcolor{black}{Comparison of the cooling power of the chillers and the TES resulting from the application of the Economic and the Energetic optimization strategies.}}
	\label{fig:Comparacion_Potencias_Econ_Ener}	
\end{figure}

A comparison between both MPC controllers in terms of thermal power is shown on Figure \ref{fig:Comparacion_Potencias_Econ_Ener}. \textcolor{black}{The economic MPC is configured with tariff A of Table \ref{tab:ElectricityRates}. In the bottom graph of this figure it can be seen that the peak thermal power, transferred and absorbed (positive and negative) by the TES system, is greater in the case of the Economic MPC for almost all diary cycles. This result suggest that the economic MPC may store and absorb more energy in the TES than the energetic one in order to accomplish with the objective of minimizing the cost.}

\begin{figure}[h!]
	\centering
	\includegraphics[width=12cm]{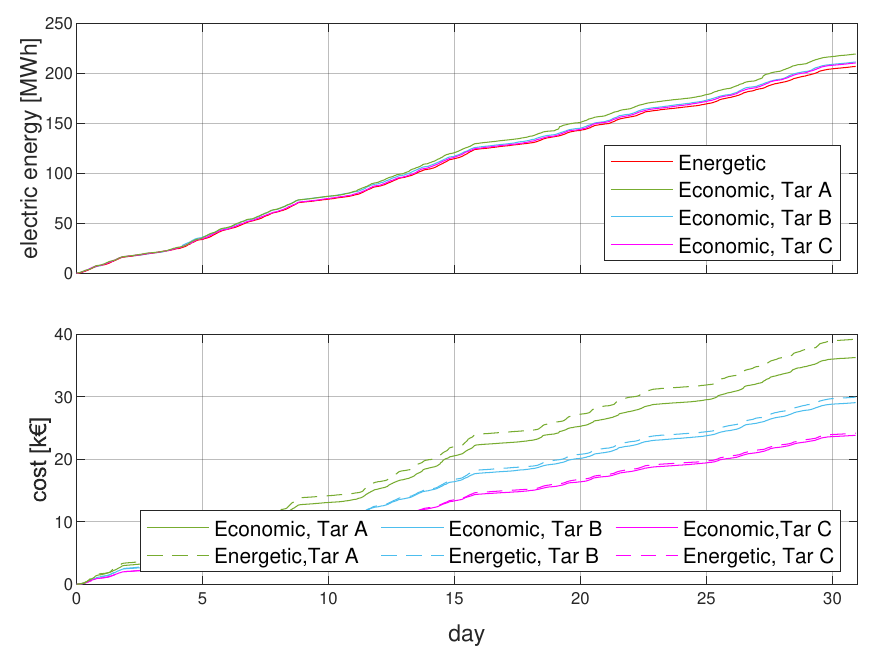}
	\caption{Economic versus non-economic MPC optimization with different electricity tariffs.}
	\label{fig:Comparacion}	
\end{figure}

\textcolor{black}{Tables \ref{tab:SummaryEnergetic}-\ref{tab:Summary} summarizes the results. The electric energy consumption of the chillers and its associated costs are shown in Tables \ref{tab:SummaryEnergetic} and \ref{tab:SummaryEconomic} for the Energetic MPC and the Economic MPC respectively. The second column of Table \ref{tab:Summary} represents the percentage cost savings that occur when using an economic optimizer versus an energetic one while the third column represents the percentage increment in consumed electric energy.} It can be seen how the Economic MPC controller reduces the total cost for all tariffs while the Energetic is the one with less total electric energy consumption, as expected. The data in these tables also show that the highest the price difference between electricity tariff periods P1 and P6, the highest are, not only the costs savings, but also the total consumed electric energy. This effect can also be seen in Figure \ref{fig:Comparacion}, in which the time evolution of the electric energy consumption of the chillers and its associated cost is shown for the two controllers and the three considered electricity rates. \textcolor{black}{The figure shows that the consumed electric energy trajectory of the energetic MPC is always below the economic MPC trajectories, and that the cost trajectory of the energetic MPC is always over the respective economic MPC trajectory for the three tariffs.}

\begin{table}[h!]
	\caption{\textcolor{black}{Energetic MPC electric energy consumption and costs in July (high electric season).} }
	\begin{center}
		\begin{tabular}{c|c|ccc}
			\multicolumn{1}{c}{} & \multicolumn{1}{c}{} & \multicolumn{1}{c}{Tar A} & \multicolumn{1}{c}{Tar B} & \multicolumn{1}{c}{Tar C} \\[1mm] 
			chiller & MWh & k\euro & k\euro & k\euro \\ \hline
			1 RTAC 400  & 76.826 & 14.888 & 11.271 & 9.074   \\
			2 RTAC 300  & 61.267 & 11.399 & 8.788  & 7.106   \\
			3 RTAC 250  & 46.320 & 8.805  & 6.716  & 5.419   \\
			4 RTAA 125  & 22.603 & 4.127  & 3.180  & 2.581   \\ \hline
			TOTAL & 207.017 & 39.220 & 29.957 & 24.181  \\		
		\end{tabular}
		\label{tab:SummaryEnergetic}
	\end{center}
\end{table}

\begin{table}[h!]
	\caption{\textcolor{black}{Economic MPC electric energy consumption and costs in July (high electric season)}. }
	\begin{center}
		\begin{tabular}{c|cc|cc|cc}
			\multicolumn{1}{c}{} & \multicolumn{2}{c}{Tar A} & \multicolumn{2}{c}{Tar B} & \multicolumn{2}{c}{Tar C} \\[1mm] 
			chiller & MWh & k\euro & MWh & k\euro & MWh & k\euro \\ \hline
			1 RTAC 400 & 78.990 & 13.546 & 78.915 & 10.956 &80.192 & 9.101   \\
			2 RTAC 300 & 64.340 & 10.595 & 61.031 & 8.468 &60.344 & 6.888 \\
			3 RTAC 250 & 51.089 &  8.273 & 47.743 & 6.543 &47.080 & 5.313  \\
			4 RTAA 125 & 25.003 & 3.883 & 23.780  & 3.107 &22.860 & 2.543   \\ \hline
			TOTAL & 219.423 & 36.298 & 211.472 & 29.075 & 210.477 & 23.846 \\		
		\end{tabular}
		\label{tab:SummaryEconomic}
	\end{center}
\end{table}

\begin{table}[h!]
	\caption{\textcolor{black}{Economic versus Energetic optimization strategy comparison summary for July.}}
	\begin{center}
		\begin{tabular}{ccc}
			tariff & cost saving [\%]& consumed electric energy increment [\%]\\ \hline
			Tar A & 7.45 & 5.99 \\
			Tar B & 2.94 & 2.15 \\
			Tar C & 1.38 & 1.67 \\
		\end{tabular}
		\label{tab:Summary}
	\end{center}
\end{table}

\FloatBarrier

\subsection{\textcolor{black}{Medium electric season results}}
{\color{black}The input profiles of the simulation for the month of September are shown in Figure \ref{fig:TrayectoriasEntradaSimulacion_Septiembre}. According to Table \ref{tab:ElectricSeasons}, the three active time periods in this month are P3, P4 and P6. Please notice that tariffs rates A and B only differentiates on the value of P1, which does not apply in medium electric season (see Table \ref{tab:ElectricityRates}), therefore only tariffs B and C will be used in this case.

\begin{figure}[h!]
	\centering
	\includegraphics[width=12cm]{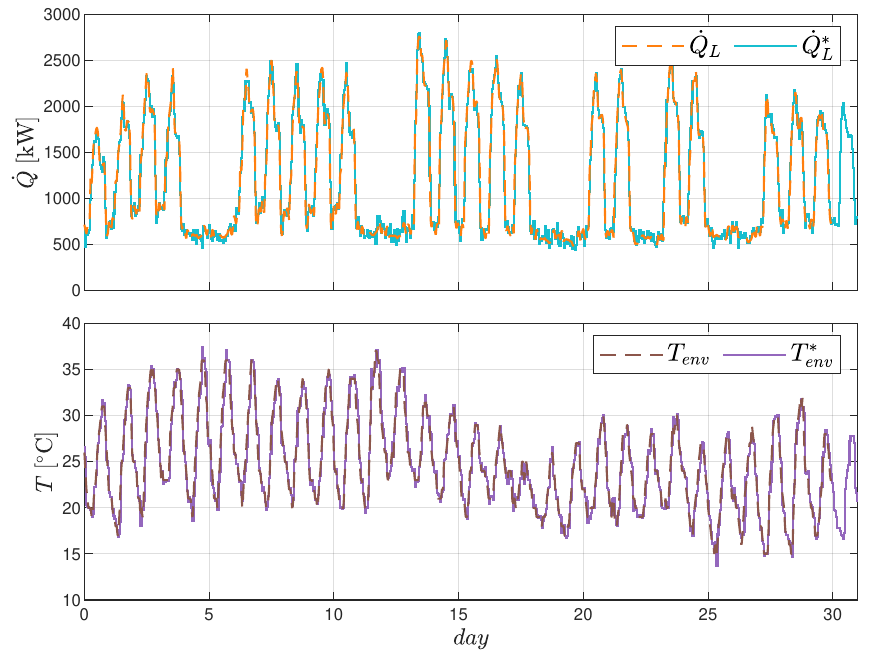}
	\caption{Simulation refrigeration load and environment temperature input profiles.}
	\label{fig:TrayectoriasEntradaSimulacion_Septiembre}	
\end{figure}

The electric energy consumption and associated costs are summarized Table \ref{tab:SummaryEnergetic_Septiembre} for the energetic optimization case, while Table \ref{tab:SummaryEconomic_Septiembre} shows the results for the economic optimization case. \textcolor{black}{Although the difference between the highest prices and the lowest prices of the tariffs are smaller than in the high electric season case, the optimizations with economic and energetic criteria still show the behaviour described in the high electric season case, that is, the economic MPC achieves lower economic costs at the cost of higher energy consumption whatever the electricity tariff is.}

}

\begin{table}[h!]
\color{black}{
	\caption{Energetic MPC electric energy consumption and costs in September (medium electric season). }
	\begin{center}
		\begin{tabular}{c|c|cc}
			\multicolumn{1}{c}{} & \multicolumn{1}{c}{} & \multicolumn{1}{c}{Tar B} & \multicolumn{1}{c}{Tar C} \\[1mm]
			chiller & MWh & k\euro & k\euro  \\ \hline
			1 RTAC 400   & 50.730 & 5.929 & 5.013    \\
			2 RTAC 300   & 43.624 & 5.056 & 4.283    \\
			3 RTAC 250   & 30.101 & 3.510  & 2.969   \\
			4 RTAA 125   & 16.727 & 1.925  & 1.633    \\ \hline
			TOTAL & 141.184 & 16.421 & 13.900  \\		
		\end{tabular}
		\label{tab:SummaryEnergetic_Septiembre}
	\end{center}
	}
\end{table}

\begin{table}[h!]
\color{black}{
	\caption{Economic MPC electric energy consumption and costs in September (medium electric season).}
	\begin{center}
		\begin{tabular}{c|cc|cc}
			\multicolumn{1}{c}{} & \multicolumn{2}{c}{Tar B} & \multicolumn{2}{c}{Tar C} \\[1mm] 
			chiller & MWh & k\euro & MWh & k\euro \\ \hline
			1 RTAC 400   & 51.433 & 5.878 & 49.565 & 4.840 \\
			2 RTAC 300   & 42.918 & 4.859 & 44.558 & 4.321 \\
			3 RTAC 250   & 30.821 & 3.501 & 30.594 & 2.965 \\
			4 RTAA 125   & 16.637 & 1.874 & 17.223  & 1.663 \\ \hline
			TOTAL & 141.811 & 16.113 & 141.941 & 13.790 \\		
		\end{tabular}
		\label{tab:SummaryEconomic_Septiembre}
	\end{center}
	}
\end{table} 

\begin{table}[h!]
\color{black}{
	\caption{\textcolor{black}{Economic versus Energetic optimization strategy comparison summary for September.}}
	\begin{center}
		\begin{tabular}{ccc}
			tariff & cost saving [\%]& consumed electric energy increment [\%]\\ \hline
			Tar B & 1.91 & 0.44 \\
			Tar C & 0.79 & 0.54 \\
		\end{tabular}
		\label{tab:SummaryMedium}
	\end{center}
	}
\end{table} 

\FloatBarrier

\subsection{\textcolor{black}{Low electric season  results}}

{\color{black}The input profiles of the simulation for the month of May are shown in Figure \ref{fig:TrayectoriasEntradaSimulacion_Mayo}. The three active time periods in this case are P4, P5 and P6. As in the medium electric season case, the price P1 does not apply and only tariffs B and C will be used in this case.

\begin{figure}[h!]
	\centering
	\includegraphics[width=12cm]{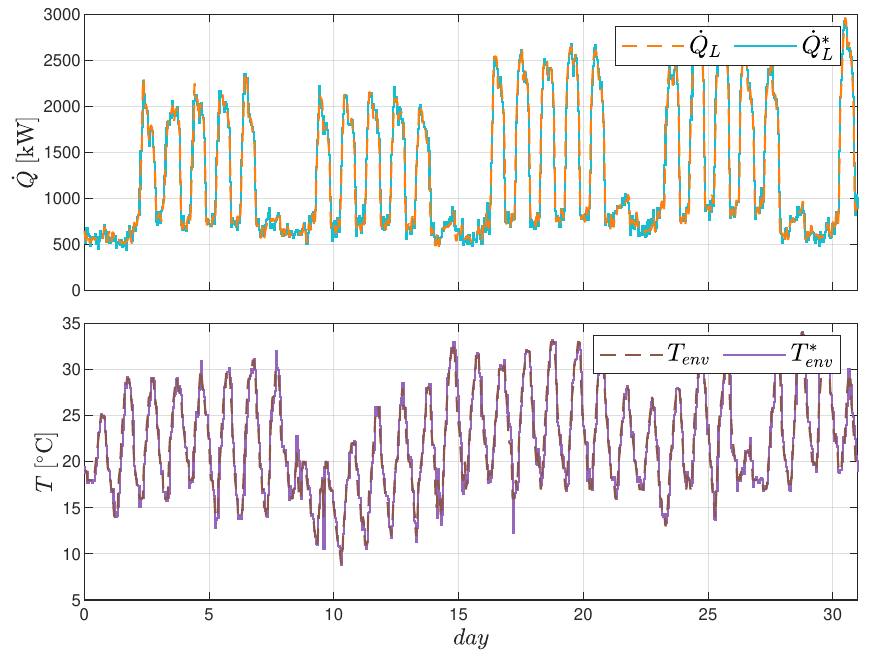}
	\caption{Simulation refrigeration load and environment temperature input profiles.}
	\label{fig:TrayectoriasEntradaSimulacion_Mayo}	
\end{figure}

The electric energy consumption and associated costs are summarized Table \ref{tab:SummaryEnergetic_May} for the energetic optimization case, while Table \ref{tab:SummaryEconomic_May} shows the results for the economic optimization case.

In the low electric season the differences between prices P4 and P6 is even lower than in the medium season, and therefore, the results of the economic and the energetic optimizations become very similar. The cost and energy savings start to be very low with the low electric season prices, and even the trend breaks in the case of tariff C (see Table \ref{tab:SummaryLow}), where the economic optimization produces a sightly higher cost than the economic one.

}

\begin{table}[h!]
\color{black}{
	\caption{Energetic MPC electric energy consumption and costs in May (low electric season). }
	\begin{center}
		\begin{tabular}{c|c|cc}
			\multicolumn{1}{c}{} & \multicolumn{1}{c}{} & \multicolumn{1}{c}{Tar B} & \multicolumn{1}{c}{Tar C} \\[1mm] 
			chiller & MWh & k\euro & k\euro  \\ \hline
			1 RTAC 400   & 55.317 & 5.856 & 5.177    \\
			2 RTAC 300   & 47.578 & 5.005 & 4.423    \\
			3 RTAC 250   & 32.993 & 3.485  & 3.076   \\
			4 RTAA 125   & 17.485 & 1.833 & 1.620    \\ \hline
			TOTAL & 153.375 & 16.181 & 14.296  \\		
		\end{tabular}
		\label{tab:SummaryEnergetic_May}
	\end{center}
	}
\end{table}

\begin{table}[h!]
\color{black}{
	\caption{Economic MPC electric energy consumption and costs in May (low electric season).}
	\begin{center}
		\begin{tabular}{c|cc|cc}
			\multicolumn{1}{c}{} & \multicolumn{2}{c}{Tar B} & \multicolumn{2}{c}{Tar C} \\[1mm] 
			chiller & MWh & k\euro & MWh & k\euro \\ \hline
			1 RTAC 400   & 57.332 & 6.003 & 54.719 & 5.101 \\
			2 RTAC 300   & 45.845 & 4.795 & 47.749 & 4.415 \\
			3 RTAC 250   & 33.131 &  3.453 & 34.140 & 3.163 \\
			4 RTAA 125   & 18.111 & 1.884 & 17.877  & 1.649 \\ \hline
			TOTAL & 154.421 & 16.138 & 154.486 & 14.329 \\		
		\end{tabular}
		\label{tab:SummaryEconomic_May}
	\end{center}
	}
\end{table} 

\begin{table}[h!]
\color{black}{
	\caption{\textcolor{black}{Economic versus Energetic optimization strategy comparison summary for May.}}
	\begin{center}
		\begin{tabular}{ccc}
			tariff & cost saving [\%]& consumed electric energy increment [\%]\\ \hline
			Tar B & 0.266 & 0.682 \\
			Tar C & -0.231 & 0.724 \\
		\end{tabular}
		\label{tab:SummaryLow}
	\end{center}
	}
\end{table}

\FloatBarrier

%%%%%%%%%%%%%%%%%%%%%%%%%%%%%%%%%%%%%%%%%%%%%%%%%%%%%%%%%%%%%%%%%%%%
\section{Conclusions} \label{Sec_7.Conclusiones}

\textcolor{black}{
The comparison between the energetic optimization objective and the economic one in MPC control of cooling plants with TES is relevant for ecological concerns. The TES system helps to reduce carbon emissions, however,  depending on the objectives of the MPC, the energy savings  will be different. The electricity tariffs are set by the Electric System Operator matching prices to electrical load in order to dissuade the consumption with the final goal of not overloading the electrical system. This decision apparently goes inline with a \textit{green} operation of the grid. However, an economically optimized cooling plant will use more electrical energy when the difference between the most expensive period and the cheapest one is higher. In this way  the economical costs are minimized, but not the energetic consumption. The results are dependent on the electric season and the available tariffs. In particular, for the high electric season and considering a representative tariff, the results show that an increment of about 2.15\% in energy consumption takes place when using the economic approach instead of the energetic one. On the other hand, a reduction in cost of 2.94\% is achieved.}

{\color{black}
The simulations show that there is a direct relation between the cost savings and energy consumption increase.  Although this consumption increase occurs during off-peak periods, relieving the electrical system, the green benefit of using a TES  is partially  lost. 

Depending on the electrical season, and therefore on the prices associated to the corresponding time periods of the tariffs, there are scenarios in which the energetic optimization beats the economic one in terms of reduction of costs. Also, it is interesting to note that, for tariff C in the low electric season, the energetic approach not only reduces consumption, but also costs. This counter-intuitive result stems from the use of a finite horizon and further supports the consideration of the energetic approach. 

As future work, a parametric analysis of the MPC strategy can be done to assess the influence of various parameters considered in the tuning of MPC which is usually a limitation of model based approaches.}

%%%%%%%%%%%%%%%%%%%%%%%%%%%%%%%%%%%%%%%%%%%%%%%%%%%%%%%%%%%%%%%%%%%%
\section*{Acknowledgements}

This research has been funded as RTI2018-101897-B-I00 Project by FEDER/Ministerio de Ciencia e Innovación - Agencia Estatal de Investigación. Spain.

%%%%%%%%%%%%%%%%%%%%%%%%%%%%%%%%%%%%%%%%%%%%%%%%%%%%%%%%%%%%%%%%%%%%
%\pagebreak
%\appendix

%\section*{References}
%\bibliography{mybibfile}

\end{document}